\documentclass[]{emulateapj}
\slugcomment{Accepted for publication in ApJ Letters}

\shorttitle{The AGN color-magnitude relation}
\shortauthors{K. Nandra et al.}

\begin{document}


\title{AEGIS: The color-magnitude relation for X-ray selected AGN}


\author{K. Nandra\altaffilmark{1}, A. Georgakakis\altaffilmark{1}, C.N.A. Willmer$^{2}$, M.C. Cooper$^{3}$, D.J. Croton$^{3}$, M. Davis$^{3}$, S.M. Faber\altaffilmark{4}, D.C. Koo\altaffilmark{4}, E.S. Laird\altaffilmark{1, 4}, J.A. Newman$^{5}$}


\altaffiltext{1}{Astrophysics Group, Imperial College London, Blackett Laboratory, Prince Consort Road,
London SW7 2AZ, United Kingdom}
\altaffiltext{2}{Steward Observatory, University of Arizona, Tucson, AZ 85721}
\altaffiltext{3}{Astronomy Department, University of California,  Berkeley, CA 94720}
\altaffiltext{4}{UCO/Lick Observatory, University of California, Santa Cruz, CA 95064}
\altaffiltext{5}{Hubble Fellow, Lawrence Berkeley National Laboratory, Berkeley, CA, 94720}


\begin{abstract}

We discuss the relationship between rest-frame color and optical luminosity for X-ray sources in the range $0.6<z<1.4$ selected from the {\it Chandra} survey of the Extended Groth Strip (EGS). These objects are almost exclusively active galactic nuclei (AGN). While there are a few luminous QSOs, most are relatively weak or obscured AGN whose optical colors should be dominated by host galaxy light. 
The vast majority of AGN hosts at $z\sim 1$ are luminous and red, with very few objects  fainter than $M_{B}=-20.5$ or bluer than $U-B=0.6$. This places the AGN in a distinct region of color-magnitude space,  on the ``red sequence'' or at the top of the ``blue cloud'', with many in between these two modes in galaxy color. A key stage in the evolution of massive galaxies is when star formation is quenched, resulting in a migration from the blue cloud to the red sequence.  Our results are consistent with scenarios in which AGN either cause or maintain this quenching. The large numbers of red sequence AGN imply that strong, ongoing star formation is not a necessary ingredient for AGN activity, as black hole accretion appears often to persist after star formation has been terminated. 
\end{abstract}




\section{Introduction}

Many studies have shown connections between rapid star formation and AGN activity in galaxies at high redshift (e.g. Page et al. 2001; Alexander et al. 2005). At $z<1$, the star formation and black--hole accretion activity in the universe both declined in a remarkably similar manner (e.g. Lilly et al. 1995; Boyle \& Terlevich 1998; Cowie et al. 2003). At low $z$, AGN are hosted predominantly by galaxies with newly-formed spheroids (Kauffmann et al. 2003; Heckman et al. 2004). These connections presumably account for the presence of  ÒdormantÓ black holes in massive, bulge dominated galaxies in the local universe (Magorrian et al. 1998; Gebhardt et al. 2000; Ferrarese \& Merritt 2000). Hence, to fully understand the formation and evolution of galaxies one must also study the AGN which they host, and vice versa.

\begin{figure*}
\epsscale{1.0}
\epsscale{1.0}
\includegraphics[angle=270,scale=0.70]{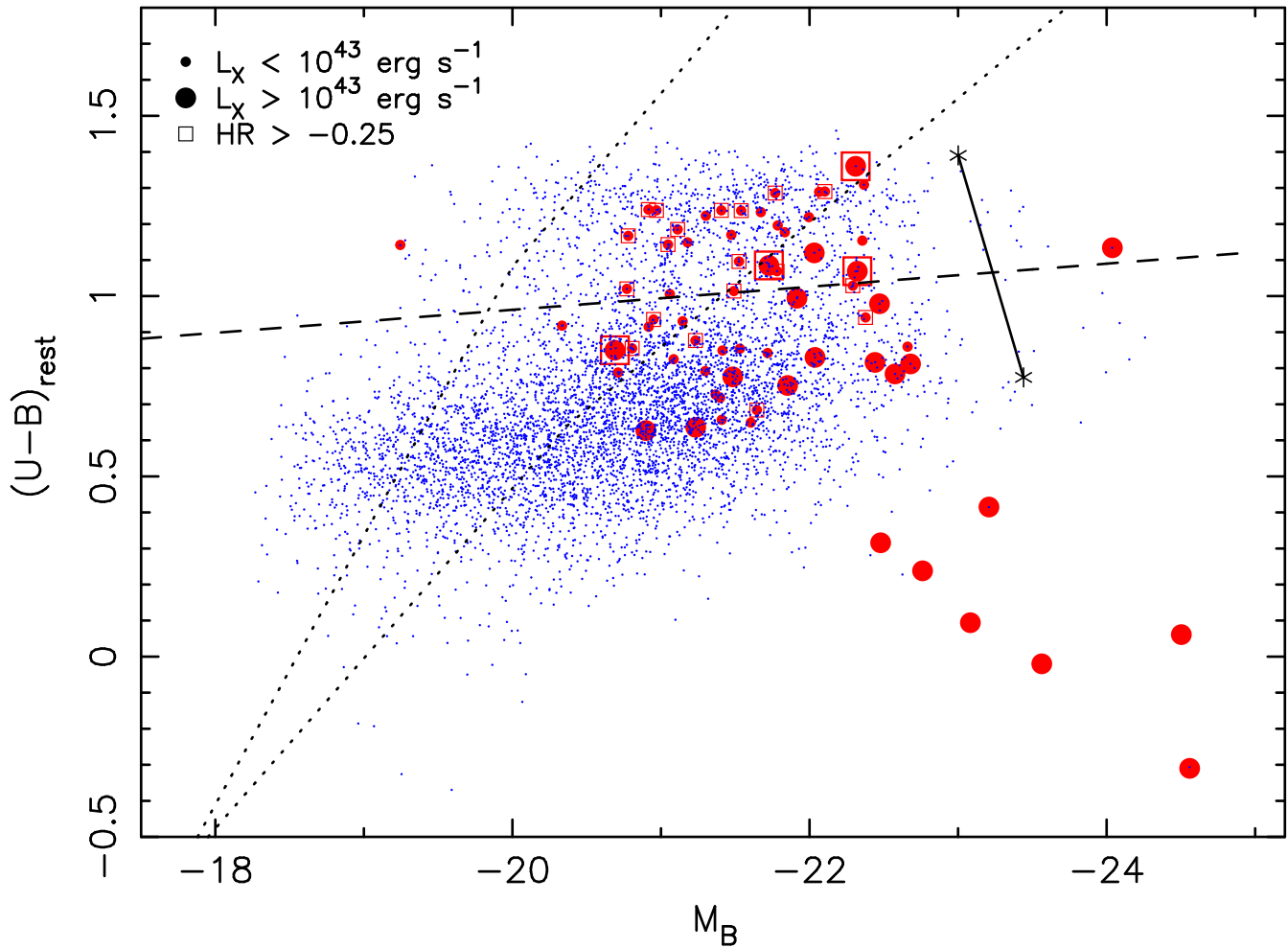}
\caption{The CMD for EGS X-ray sources. Rest-frame $U-B$ color is shown against$B$-band absolute magnitude for DEEP2 comparison galaxies (small blue dots) and X-ray sources (red circles) in the range $0.6<z<1.4$. Squares around the symbols indicate hard X-ray sources, and more luminous systems ($L_{\rm X}$(2-10 keV) $>10^{43}$erg s$^{-1}$) are plotted with larger symbols. 
The dashed line separates red and blue galaxies (Willmer et al. 2006) and dotted lines show the DEEP2 completeness limits at $z=1.0$ and $z=1.4$ (Gerke et al. 2006).  The solid line shows the effect of AGN contamination on the color: the upper star represents a pure elliptical template (Coleman, Wu \& Weedman 1980), the lower one is contaminated by a QSO contribution  (Vanden Berk et al. 2001) with half the flux of elliptical in the B-band. The great majority of X-ray sources lie in luminous ($M_{\rm B}<-20.5$), red galaxies in and around the transition region between the ``blue cloud'' of active star forming galaxies and the ``red sequence'' of passively evolving bulges. This finding is consistent with AGN activity being associated with the process which quenches star formation in massive galaxies, causing the  migration of blue galaxies to the red sequence. }
\label{fig:cmd}
\end{figure*}

Great progress has been made in understanding the formation of massive galaxies with the recognition that  the distribution of their rest-frame colors is bimodal (e.g., Strateva et al. 2001; Im et al. 2002; Baldry et al. 2004; Bell et al. 2004; Weiner et al. 2005). The ``red sequence''  consists mainly of massive, passively evolving early type galaxies (E/S0s), while the majority of galaxies show bluer colors due to ongoing star formation. A relatively narrow valley separates galaxies in color space. The emerging consensus is that mergers fuel intense star formation at high redshift, which builds massive bulges. At some point, and by a mechanism which is hitherto unclear, star formation is quenched and they migrate from the blue cloud to the red sequence (Bell et al. 2004; Faber et al. 2006). The red galaxies then evolve passively, without further significant star formation.

A variety of mechanisms have been proposed to account for quenching (e.g. Moore et al. 1996; Cox et al. 2004; Cattaneo et al. 2006), but much recent work has suggested that AGN may be responsible for the moderation of star--formation activity. Winds driven by violent black hole activity, perhaps driven by mergers, can sweep up gas from the galaxy, stopping star formation. This can plausibly provide the necessary link between black hole accretion and galaxy evolution (Silk \& Rees 1998; King 2003; Murray et al. 2005; Hopkins et al. 2005). Alternatively, galaxies may enter massive static haloes in which gas cannot cool efficiently (Dekel \& Birnboim 2006) and star formation is suppressed. Thereafter, low-level AGN at the centers of groups and clusters can inhibit further gaseous cooling and infall (Croton et al. 2006). In these scenarios one would predict AGN to occupy a distinct region of the color-magnitude diagram (CMD) which defines and delineates the bimodality of galaxy colors. In this {\it Letter}, we use the AEGIS  data (Davis et al. 2006) to test this idea. 

We use AB magnitudes throughout and adopt a flat cosmology with $\Omega_{\rm \Lambda}$=0.7 and $h=0.7$.

\section{Observations and analysis}

Deep X-ray surveys are currently the most efficient and reliable way to select AGN, with many being indistinguishable from normal galaxies in the optical or mid-IR (e.g., Mushotzky et al. 2000; Barmby et al. 2006). The central engine may be obscured, weak, or both, but if the galaxy dominates the optical light, this clearly greatly facilitates the study of the AGN host. We have used the {\it Chandra} sample described in Georgakakis et al. (2006a) to select AGN in the EGS. The data were analysed using the methods described in Nandra et al. (2005).  {\it Chandra} sources were matched to the DEEP2 catalogues as described in Georgakakis et al. (2006b). We include a small number of X-ray source redshifts from DEEP1, CFRS, and SDSS, which satisfy the $R\leq24.1$ photometric limit of DEEP2. We restrict our study to spectroscopic galaxies in the range $0.6<z<1.4$, where the redshift success rate is excellent. Four objects were excluded from the sample due to suspect redshifts or K-corrections. The final sample consists of 68 X-ray sources, with a comparison sample of 5864 DEEP2 galaxies. 

Rest-frame $U-B$ values and B-band absolute magnitudes ($M_{\rm B}$) have been calculated using the method and K-corrections of Willmer et al. (2006). Rest-frame 2-10 keV X-ray luminosities $L_{\rm X}$  have been calculated from the 2-10 keV observed-frame flux using a typical intrinsic AGN spectrum of $\Gamma=1.9$ (Nandra \& Pounds 1994). This effectively produces absorption-corrected fluxes for sources with column densities $N_{\rm \rm H}< 10^{23}$~cm$^{-2}$ at $z\sim1$. In addition, the hard ($\sim 4-20$ keV rest-frame) bandpass ensures that the contribution of hot gas to the X-ray flux is negligible. Where the 2-10 keV flux was undetermined, the 0.5-10 keV flux was used to calculate $L_{\rm X}$ (2-10 keV).

%% In this section, we use  the \subsection command to set off
%% a subsection.  \footnote is used to insert a footnote to the text

%% This section also includes several of the displayed math environments
%% mentioned in the Author Guide.

\section{The AGN color-magnitude relation}

The color-magnitude diagram (CMD) for the X-ray sources and comparison galaxies is shown in Fig.~\ref{fig:cmd}. The latter show the well-established bi-modality of colors at this redshift. It is clear that X-ray sources are not randomly distributed over the same region as the optical galaxies. Firstly, there are a number of very luminous, blue sources that lie outside the field galaxy color-magnitude relations. They are luminous and soft in the X-ray and and inspection of the HST images confirms that they are dominated in the optical by a central, blue point source, i.e. they are QSOs.

\begin{deluxetable}{lcccccc}
\tabletypesize{\scriptsize}
\tablecaption{AGN and field galaxies in various regions of the CMD.}
\tablewidth{0pt}
\tablehead{
\colhead{Sample$^{a}$} &  
\colhead{$N_{\rm X}^{b}$} & 
\colhead{$F_{\rm X}^{c}$} & 
\colhead{$N_{\rm Gal}^{d}$} & 
\colhead{$F_{Gal}^{e}$} &
\colhead{$p_{\rm field}^{f}$} &
\colhead{$F_{\rm field}^{g}$} 
}
\startdata
Red sequence		  	& 26 & 46\% 	& 753 	& 21.6\% 	&  $3 \times 10^{-4}$ & 3.5\% \\
Valley				& 7   & 13\% 	& 218	& 6.3\% 	&  $7 \times 10^{-2}$  & 3.2\% \\
Blue cloud 			& 23 & 41\% 	& 2513 	& 72.1\% 	&  $3\times  10^{-3}$ & 0.9\% \\
\enddata
\tablenotetext{a}{See text for definition of these samples. The analysis is restricted to $-22.5< M_{\rm B}<-20.5$}
\tablenotetext{b}{Number of X-ray sources in the region}
\tablenotetext{c}{Percentage of X-ray source population out of 56 in $M_{\rm B}$ range}
\tablenotetext{d}{Number of comparison DEEP2 galaxies }
\tablenotetext{e}{Fraction of galaxy population in the region (total 3484)}
\tablenotetext{f}{Probability of observing $N_{\rm X}$ based on $F_{\rm gal}$}
\tablenotetext{g}{Percentage of X-ray sources among the control population}

\end{deluxetable}

Even when these are discounted, AGN still reside almost exclusively in luminous galaxies, with only 2 of the 68 X-ray detected AGN host galaxies fainter than  $M_{\rm B}>-20.5$. The X-ray sources are also overwhelmingly among the redder galaxies, confirming the work of Barger et al. (2003), with 90\% of X-ray sources having $U-B>0.6$ and 46\% $U-B>1$. The probabilities of obtaining these fractions by chance are negligible. 

It is evident from the CMD that most AGN reside on the red sequence, at the top of the blue cloud, and in the valley between them. We have divided the sample into red and blue galaxies according to the criterion of Willmer et al. (2006), and a range $\pm 0.05$ in $U-B$ about the relation delineating the valley. We further restrict the sample to $-22.5< M_{\rm B}<-20.5$ which avoids most of the QSOs, as well as  the low--luminosity host galaxies where AGN are absent.  We find significantly enhanced AGN activity on the red sequence and in the valley, even when restricting the $M_{\rm B}$ range. The fraction of galaxies which are X-ray sources in the red sequence, valley and blue cloud are 3.4, 3.2 and 0.9\%. Details of the number of sources in the  various ranges are given in Table~1. 

\begin{figure}
\epsscale{1.0}
\includegraphics[angle=270,scale=0.30]{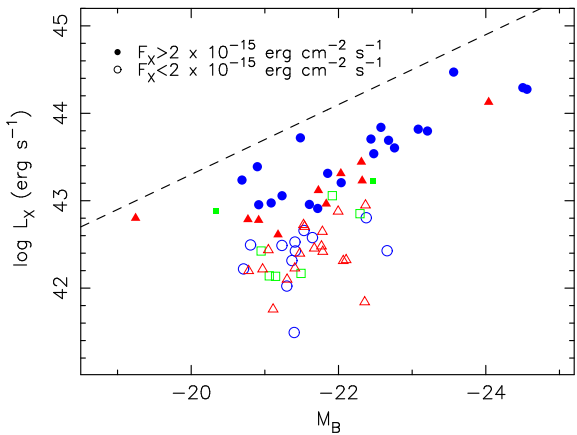}
\caption{2-10 keV X-ray luminosity versus $M_{\rm B}$. X-ray sources have been segregated by optical color (red triangles: red sequence, blue circles: blue cloud, green squares: valley). As one probes to fainter X-ray fluxes and luminosities (open symbols), X-ray sources still inhabit only the most luminous optical hosts. The two luminosities also become uncorrelated, since the AGN dominates in the X-ray and the host galaxy in the optical. The dashed line shows the slope of a linear relation between $L_{\rm X}$ and $M_{\rm B}$ to guide the eye.} 
\label{fig:mb_lx}
\end{figure}

The X-ray and optical surveys are both flux limited, so it is important to consider potential selection biases. The apparent lack of galaxies with $M_{\rm B}>-20.5$ might, e.g., be due to the X-ray flux limit. This does not seem to be the case.  Fig.~\ref{fig:mb_lx} shows the relationship between $L_{\rm X}$ and $M_{\rm B}$. At the highest luminosities ($M_{\rm B}<-22.5$), there is a trend in which the X-ray and optical luminosities of the galaxies are correlated, consistent with the light from the two bands coming from the AGN. This largely disappears at $M_{\rm B}>-22.5$, however, where the points are essentially evenly distributed. Furthermore, as one probes to lower fluxes and X-ray luminosities ($L_{\rm X}<10^{43}$~erg s$^{-1}$), X-ray sources continue to populate only luminous host galaxies, even though they could easily be detected in fainter ones.

The DEEP2 limit of $R<24.1$ also introduces a color-dependent completeness limit (Fig.~\ref{fig:cmd}). Specifically, there are relatively few red galaxies with $M_{\rm B}>-20.5$ and they will mostly be at $z<1$. Based on the fraction of AGN with $U-B>1$ and $-22.5<M_{\rm B}<-20.5$,  however, the finding of only 1 galaxy with $M_{\rm B}>-20.5$ and $U-B>1$ is still significant at $99$\% confidence. If there is differential evolution between the AGN and normal galaxies, the redshift-dependent DEEP2 completeness limit (Fig.~\ref{fig:cmd}) might introduce a bias. AGN evolution at these luminosities in the range $z=0.6-1.4$ is relatively mild (e.g., Ueda et al. 2003), and we see no significant difference in the redshift distributions of the X-ray and non X-ray samples, so this should not be a problem. Overall, biases due to optical selection effects should be small because they apply equally to the X-ray and field samples. The simplest test that verifies this is the observation that the the X-ray sources host galaxies are significantly brighter in {\it observed} magnitude, with only 7\% being within $\Delta R < 0.5$ of the DEEP2 limit compared to 33\% of the general population. 

%However, and no significant difference is seen in the reds
%there are 162 galaxies. The percentage of galaxies with an X-ray source that have MB -20 to -22 and %U-B>1 is 3.9%. Therefore we would expect 6.3 X-ray sources in that quadrant, whereas we see none. I %think this has negligible chance probability.

Another critical issue is whether the AGN significantly contaminates the host galaxy light. This is clearly the case in the luminous blue sources in Fig~\ref{fig:cmd}. Fig.~\ref{fig:mb_lx} shows, however that at low X-ray luminosities the correlation between $L_{\rm X}$ and $M_{\rm B}$ is weak. Visual inspection of the HST images of the AGN hosts (Pierce et al. 2006) shows that about 1/3 of the galaxies lying on the red sequence and/or blue cloud have unresolved, blue nuclei. For the majority of the sample, the contamination is therefore minor, but its effect will be to reduce the number of red sequence AGN seen in the sample (see template track in Fig~\ref{fig:cmd}). Indeed, if we restrict the analysis only to hard sources, whose nuclear point sources are presumably completely obscured in the optical, the fraction of X-ray sources found on the red sequences rises from 46\% (Table~1) to 65\% (15/23 objects). We also note in passing that Fig.~\ref{fig:mb_lx} shows that very deep X-ray data are required for a full picture. For example, the majority of red sequence AGN would not have been detected in a substantially shallower X-ray survey. 

\section {Discussion}

%Much recent evidence has pointed towards the conclusion that AGN reside preferentially in massive 
%galaxies (e.g. Dunlop et al. 2003; Kauffman et al. 2003; Colbert et al. 2005). We confirm this, at least if $M_{\rm B}$ is a reasonable tracer of mass, with 95\% of X-ray detected AGN at our flux limit being more luminous than $M_{\rm B}=-20$.  Barger et al. (2003) have furthermore noted that the rest-frame colours of many X-ray source host galaxies are red, suggesting they have evolved populations. We also confirm a large fraction  of AGN hosts with red colors, e.g. 43\% have rest $U-B>1$, compared to only 17\% of all galaxies.

X-ray selected AGN at $z\sim 1$ occupy a distinct place in galaxy color-magnitude space. They are predominantly associated with luminous ($M_{\rm B}<-20.5$), red ($U-B>0.6$) hosts. The great majority  lie on the red sequence, at the top edge of the blue cloud, and the region in between, exactly where massive star-forming galaxies are able to transition into passive, early types. These findings strongly suggest a link between AGN and the process responsible for that transition, and the quenching of star formation in massive galaxies. 

Our results are broadly consistent with those found locally for SDSS AGN,  whose hosts are massive, early type galaxies with evidence for ongoing or recent star-formation (Kauffmann et al. 2003; Heckman et al. 2004). Their work concludes that the pre-requisites for an AGN are a massive black hole -- axiomatically associated with a bulge  -- and a gas supply which fuels both star formation and accretion. Our CMD at $z\sim 1$ is consistent with this picture, and is also in agreement with the large fraction of early-type morphologies seen in X-ray selected AGN (Grogin et al. 2005, Pierce et al. 2006). Sanchez et al. (2004) also find early type morphologies for an optically-selected AGN sample. Their CMD differs from ours, in that it shows a wide range of host galaxy colors. This difference may well be due to the different selection methods, with the Sanchez et al. sample consisting exclusively of type 1s. Their conclusion is that the AGN are generally bluer for their morphological type due to star formation. 

One way of interpreting these results is to suggest that star formation promotes AGN activity, e.g., via starburst winds fuelling the central black hole. The fact that we see so many X-ray selected AGN on the red sequence at $z\sim 1$ shows that AGN persist {\it after} star formation has largely ended, meaning that strong, ongoing star formation is not a pre-requisite for AGN activity. In fact, because the AGN contribute blue light to the host galaxies, it may be that we are significantly underestimating the number of galaxies on the red sequence, due to contamination. 

Alternatively, AGN activity might suppress star formation in massive galaxies. This idea is currently very popular, as it can in principle provide the necessary connection between black holes and bulges seen in local galaxies.  The boldest interpretation is that the AGN actually terminates star formation (Hopkins et al. 2005; Springel, Di Matteo \& Hernquist 2005).  In this ``QSO-mode'', the initial trigger is a major merger, which provides gas both for star formation and to feed the AGN. Initially the QSO is cloaked in gas and obscured, but  it eventually sweeps up the obscuring and starforming gas via powerful winds, blowing it out of the galaxy. 

The position of the X-ray sources on the CMD generally supports this scenario, in that AGN activity occurs in and around the transition. There may be some differences in detail, however. The QSO-mode is merger-driven, but morphological disturbances in X-ray selected AGN are not clearly seen (Grogin et al. 2005; Pierce et al. 2006). One might also predict a very large fraction of the ``valley" galaxies to be AGN, if they are in the process of quenching. While we do find many AGN in the valley, they are far from universal and, e.g., are no more common than on the red sequence.  Finally, the Hopkins et al. (2005) model explicitly predicts that  AGN should be obscure, and thus have a hard X-ray spectrum before red-sequence migration, at which point they blow the obscuring, star-forming gas out of the galaxy and become soft. We find many soft X-ray sources among the blue galaxies, and many hard X-ray sources on the red sequence (Fig.~\ref{fig:cmd}), disfavouring this picture and indeed any in which star-forming gas is the primary source of obscuration.  The requirement is now for quantitative comparisons between our data and the simulations on which Hopkins et al. based their predictions. 

Another suggestion is the so-called ``radio-mode'' of Croton et al. (2006). Here, AGN heating is invoked to prevent hot gas in group and cluster environments cooling and falling into the galaxy to form new stars. This model predicts relatively low-level AGN activity in massive galaxies in dense environments. Much previous work has shown that AGN hosts are massive (e.g. Kauffman et al. 2003; Colbert et al. 2005), and our data support this conclusion specifically for our $z\sim 1$ X-ray selected sample. The complementary work of Georgakakis et al. (2006a) shows that they also reside in rich environments at this redshift. If these massive galaxies host correspondingly large black holes, these weak AGN must be radiating well below the Eddington limit, consistent with the Croton et al. (2006) picture. While it is true that the majority of AGN on the red sequence have relatively low luminosity (Fig.~\ref{fig:cmd}), it should be noted that the numbers are small and, moreover, powerful AGN would likely cause the objects to move off the red sequence due to nuclear contamination (Fig.~\ref{fig:cmd}).  Again, more quantitative comparisons are now required to determine whether the heating rates, implied AGN duty cycle and environmental measures are consistent in detail. For now, this AGN ``radio-mode'' appears to be a very promising candidate for maintaining quenching in massive galaxies.

While the initial causes of quenching remain unresolved, our results argue that AGN have an important connection to the late-stage evolution of massive galaxies.  The challenges for the full AEGIS dataset will be to determine the physical mechanism behind this, and to discern what activates the AGN in the first place. 

\acknowledgments
We thank the anonymous referee and Phil Hopkins for helpful comments. We acknowledge financial support from the Leverhulme trust (KN), PPARC (AG), Marie-Curie fellowship grant MEIF-CT-2005-025108 (AG), Chandra grant GO5-6141A (DCK), Hubble Fellowship grant HST-HF-01165.01-A (JAN), and NSF grants  AST-0071198,  AST-0071048 and AST-0507483. The W.M. Keck Observatory is a scientific partnership among Caltech, the University of California and NASA, made possible by the support of the W.M. Keck Foundation. We wish to acknowledge the very significant cultural role that the summit of Mauna Kea has within the indigenous Hawaiian community; we are fortunate to be able to conduct observations from this mountain.

\end{document}